\documentclass[12pt]{article}
\usepackage{amsmath,amssymb,amsfonts,epsfig}
\usepackage[nosort]{cite}

\makeatletter\@addtoreset{equation}{section}\makeatother

\def\bR {\mathbb{R}}
\def\bZ {\mathbb{Z}}

\input epsf

\newcommand{\figin}[2]{
\begin{figure}[t]
\centerline{\hbox{\epsffile{#1.eps}}}
\centerline{\parbox{12cm}{\caption{#2\label{#1-fig}}}}
\end{figure}}

\newcommand{\vev}[1]{{\left< {#1} \right>}}

\newcommand{\ket}[1]{{\left| {#1} \right>}}

\newcommand{\Tr}{{\rm Tr\,}}

\newcommand{\cC}{{\cal C}}

\newcommand{\cE}{{\cal E}}

\newcommand{\cJ}{{\cal J}}
\newcommand{\cL}{{\cal L}}
\newcommand{\cN}{{\cal N}}
\newcommand{\cO}{{\cal O}}
\newcommand{\cP}{{\cal P}}
\newcommand{\cS}{{\cal S}}

\newcommand{\preprint}[1]{\begin{table}[t]  
             \begin{flushright}               
             {#1}                             
             \end{flushright}                 
             \end{table}}                     
\renewcommand{\title}[1]{\vbox{\center\LARGE{#1}}\vspace{5mm}}
\renewcommand{\author}[1]{\vbox{\center#1}\vspace{5mm}}
\newcommand{\address}[1]{\vbox{\center\em#1}}
\newcommand{\email}[1]{\vbox{\center\tt#1}\vspace{5mm}}

\begin{document}

\begin{titlepage}
\preprint{
hep-th/0604124
}

\title{Small deformations of supersymmetric Wilson loops 
and open spin-chains}

\author{Nadav Drukker$^1$ and Shoichi Kawamoto$^2$}

\address{$^1$The Niels Bohr Institute, Copenhagen University\\
Blegdamsvej 17, DK-2100 Copenhagen, Denmark\\
\smallskip
$^2$Rudolf Peierls Centre for Theoretical Physics, 
University of Oxford,\\
1 Keble Road, Oxford, OX1 3NP, United Kingdom
}

\email{drukker@nbi.dk, kawamoto@thphys.ox.ac.uk}

\abstract{
We study insertions of composite operators into Wilson loops in 
$\cN=4$ supersymmetric Yang-Mills theory in four dimensions. The 
loops follow a circular or straight path and the composite insertions 
transform in the adjoint representation of the gauge group. This provides 
a gauge 
invariant way to define the correlator of non-singlet operators. Since 
the basic loop preserves an $SL(2,\bR)$ subgroup of the conformal 
group, we can assign a conformal dimension to those insertions 
and calculate the corrections to the classical 
dimension in perturbation theory. 
The calculation turns out to be very similar to that of single-trace local 
operators and may also be expressed in terms of a spin-chain. 
In this case the spin-chain is open and at one-loop order has Neumann 
boundary conditions on the type of scalar insertions that we consider. 
This system is integrable and we write the Bethe ansatz describing it. 
We compare the spectrum in the limit of large angular momentum both 
in the dilute gas approximation and the thermodynamic limit to the 
relevant string solution in the BMN limit and in the full 
$AdS_5\times S^5$ metric and find agreement.
}

\end{titlepage}

\section{Introduction}

Over the past few years great progress has been made on calculating 
the spectrum of $\cN=4$ supersymmetric Yang-Mills theory in 
four dimensions at large $N$. In the weak coupling regime the 
perturbative calculation of the conformal dimensions of operators 
was mapped to certain integrable spin-chain models. At strong 
coupling the system is described by strings propagating on 
$AdS_5\times S^5$ and one may use the integrability of the 
classical $\sigma$-model to calculate the dimensions of operators 
carrying large charges.

The simplest non-trivial scalar operators are words of arbitrary length 
composed of two complex scalar fields. Such operators are said to be in 
the $SU(2)$ sector and are the ones that were studied the most. The 
one-loop anomalous dimension was related to the spectrum of the 
spin-$1/2$ Heisenberg spin-chain \cite{Minahan:2002ve}. The two 
and three loop anomalous dimensions (as well as a conjectured 
all-loop expression) are given by the spectra of Hamiltonians of 
spin-chains with longer range interactions 
\cite{Beisert:2003tq,Beisert:2003ys,Eden:2004ua}.
Recently they were related to another type of condensed matter system, 
the Hubbard model, where all the long-range interactions are repackaged 
as local interactions for some dual variables \cite{Rej:2005qt}. 
Other local operators involving the fermions and gauge-fields 
were also studied to different degrees (see for example 
\cite{Beisert:2003yb}).

On the string theory side the integrability of the classical 
$\sigma$-model was established in \cite{Bena:2003wd}. 
Classical solutions corresponding to operators carrying large charges 
were found even before, most notably the BMN states rotating on 
$S^5$ \cite{Berenstein:2002jq} and the strings spinning in 
$AdS_5$ \cite{Gubser:2002tv}. Many more solutions corresponding 
to states carrying different collections of charges were found since 
then and in many cases the spectra as calculated from string 
theory agrees with the perturbative result%
\footnote{For more details on the calculations of the spectrum of 
the dilatation operator see the reviews 
\cite{Tseytlin:2003ii,Beisert:2004ry,Plefka:2005bk}.
}.

Another interesting class of gauge invariant operators are Wilson loops, 
the trace of the holonomy around a closed contour. Within the 
$AdS$/CFT correspondence \cite{Maldacena:1997re} they are 
evaluated in terms of fundamental strings that extend to the boundary 
of space \cite{Rey,Maldacena-wl} (they may also be described in 
certain cases by D3-branes and D5-branes 
\cite{Drukker:2005kx,Gomis:2006sb,Yamaguchi:2006tq}). 
Over the years the surfaces corresponding to several different Wilson 
loops were found 
\cite{Berenstein:1998ij,Drukker:1999zq,Zarembo:1999bu,
Zarembo:2002an,Tseytlin:2002tr,Mikhailov:2002ya,
Dymarsky:2006ve}. 
Still much less is know about those observables compared to local 
operators.

In \cite{Drukker:2005cu} the integrability of the classical string 
$\sigma$-model in $AdS_5\times S^5$ was used to organize 
the calculation of the surfaces associated with Wilson loop observables. 
In that paper certain Wilson loops with periodic shapes and scalar 
couplings were evaluated by imposing a similar periodic ansatz on the 
string solution. That reduced the $\sigma$-model to a finite dimensional 
integrable system that was then solved classically, allowing to evaluate 
those Wilson loops at strong coupling. But so far similar techniques 
have not been found to calculate the expectation values of Wilson loops 
or their 2-point functions directly in the gauge theory. In this paper we 
find a spin-chain that describes the expectation values of some Wilson 
loop observables.

As we shall review, some very special Wilson loop operators (a single 
straight line or a circle with constant coupling to one of the scalars) 
preserve a subgroup of the conformal group in four dimensions which 
includes an $SL(2,\bR)$ factor. This is the group of rigid conformal 
transformations in one dimension and we will classify deformations 
of the symmetric Wilson loop by representations of this group.

This is in direct analogy to local operators, they are deformations 
of the conformally invariant vacuum and may therefore be classified 
in representations of the conformal group. Instead of the vacuum we 
are starting with the background which includes an $SL(2,\bR)$-invariant 
Wilson loop and will classify its deformations in terms of 
representations of that group.

In \cite{Drukker:2005af} we established this symmetry and 
proposed it as a way of classifying Wilson loops. In most of the 
current paper we take a slightly different philosophy and instead of 
studying the representations of the Wilson loops, we isolate the 
representations (or dimensions) of the local insertions into the loop. We 
make some comments on the representations of the Wilson loop itself 
towards the end of the paper.

A general Wilson loop operator will be very different from the symmetric 
one, much like a general state in the gauge theory is not necessarily 
created by one local operator. But to simplify the general case one 
usually starts with a single local operator, likewise we will study a local 
deformation of the Wilson loop.

The type of deformation of the loop on which we will concentrate is an 
insertion of a word made up of some of the six scalars (we will restrict 
further to two complex combinations of them) all at one point along the 
loop. The insertion will transform in the adjoint representation of the 
gauge group and is made gauge invariant by the Wilson loop on which 
it is sitting. More general insertions will include the fermionic fields 
as well as field-strengths and covariant derivatives. One can also write 
those insertions in terms of functional derivatives of the basic loop 
with respect to its path and other couplings.

The standard way of calculating the conformal dimension of a local 
operator is by evaluating the two-point function with another local 
operator. Mimicking that, we will consider the insertion of another 
``probe'' operator into the Wilson loop. Thus we can regard the 
expectation value of the Wilson loop with the insertion of two 
operators as a gauge invariant definition for the two-point function 
of adjoint operators.

In Section~\ref{sec-prelim} we will review \cite{Drukker:2005af} 
and establish the notion of the conformal dimension of a local 
deformation of the circular and straight Wilson loops.

Having set up the framework we proceed to calculate the dimensions 
of those insertions by directly evaluating at one-loop in perturbation
theory their two-point functions (i.e. the vacuum expectation value 
of the Wilson loop with the two insertions). We do this in the planar 
approximation, valid for large~$N$. In our case, the difference from 
single-trace operators is that the insertions are not cyclical, each has a 
beginning and an end and in the planar approximation the order has 
to be kept. At tree-level this implies that the two point function vanishes 
unless the two words are exactly identical (with the reversed order).

At one-loop the planar diagrams allow interactions between nearest 
neighbors, and those interactions are exactly the same as between the 
letters in a single-trace local operator. The exception are the outermost 
fields, which cannot interact with each-other by planar graphs. 
Instead they interact with the rest of the Wilson loop. Just from 
those considerations about the possible planar graphs we 
immediately see that the problem of calculating the one-loop mixing 
matrix of those insertions can be mapped to the spectral problem 
of a Hamiltonian of an {\em open} spin-chain. The interaction 
between the outermost fields and the Wilson loop provides the boundary 
terms for the open-chain Hamiltonian.

For the type of insertions we will consider this interaction will 
not depend on the flavor indices, and will be a constant. We find 
the regular $SU(2)$ open spin-chain with Neumann boundary 
conditions.

This system is integrable and may be solved in terms of the Bethe 
ansatz. Anybody familiar with those techniques used to calculate the 
dimension of local operators would immediately recognize the 
solution, but we will present it in detail in Section~\ref{sec-Bethe}. 
There we will derive the spectrum in the dilute gas approximation 
and in Section~\ref{sec-thermo} we calculate it in the thermodynamic 
limit.

Next, in Section~\ref{sec-strings} we study the same system in 
string theory, by finding the relevant classical solution to the string 
equations of motion on $AdS_5\times S^5$. We start with the 
Wilson loop containing insertions of only one of the two complex 
scalars, which preserves $1/4$ supersymmetry and is an open-string 
analogue of the BMN vacuum \cite{Berenstein:2002jq}. To 
preserve the maximal possible symmetry, one insertion is mapped to 
past infinity in global Lorentzian $AdS$-space and the other 
insertion to future infinity. The boundary gauge theory for this 
system lives on $\bR\times S^3$, and our Wilson loop will go 
from past infinity to the future and then back along antipodal points 
on the $S^3$. We present the string solution for these boundary 
conditions and calculate the angular momentum carried by it.

To study the deformations of the basic solution we will 
take the BMN limit, concentrating near the center of the geometry, 
where the string solution is mapped to an infinite surface in the 
pp-wave geometry. Using a cutoff on the size of this string, we 
can calculate its spectrum of fluctuations and compare it to the solution 
of the Bethe equations. We find complete agreement with the leading 
order result.

We go further and study the system with two angular momenta, 
which corresponds to a large number of both of the complex 
fields. The equations we find are identical to those of certain folded 
string solutions, but again describing open strings. Instead of the 
string folding on itself it extends infinitely to the boundary of 
$AdS_5$ and a fixed point on $S^5$.

In Section~\ref{sec-local} we go back to studying the question 
suggested in \cite{Drukker:2005af}, of evaluating the dimension of 
a Wilson loop (rather than of the insertion into a Wilson loop). 
The Wilson loop we consider will be circular (or straight) and have 
an insertion of a single adjoint operator, and as argued there it may 
be organized in irreducible representations of $SL(2,\bR)$. To study 
them at the one-loop level in perturbation theory we consider the 
correlator of such a Wilson loop with a single-trace local operator or 
with another Wilson loop.

In order to have control over the functional form of the two-point 
function we need the local operator to coincide with a point along 
the Wilson loop (or for two loops, they will have to coincide). Then 
we can again interpret any divergence as corrections to the conformal 
dimension (i.e. $SL(2,\bR)$ representation) of the Wilson loop.

Now there is a single insertion in the loop, which is traced over. So 
in terms of spin-chains this will correspond to a closed chain, similar 
to local operators. Still the system does not posses cyclical symmetry, 
at the planar level only the outermost fields in the insertion can 
interact with the Wilson loop. Those interaction graphs will introduce 
extra terms in the spin-chain Hamiltonian localized at a fixed 
position along the chain.

From this simple analysis of the possible planar diagrams we find a 
closed spin-chain with a marked point specifying where the word 
starts and ends. In the planar approximation, all interactions between 
the Wilson loop and the insertion will be around this marked point. 
At the one-loop level, which we calculate explicitly, this interaction 
is a flavor independent constant, giving a constant shift of the 
dimension of the loop compared with the single-trace local operator 
made out of the same word.

We conclude with a discussion of the meaning of the calculations we 
have performed. We also present some open questions and 
generalizations that will be left for future work.

In Appendix~\ref{sec-calc} we give the details of the calculation 
of the Feynman graphs involving interactions of the insertions with 
the Wilson loop.

\section{Preliminaries}
\label{sec-prelim}

In this paper we are considering Wilson loop operators in $\cN=4$ 
supersymmetric Yang-Mills theory
\begin{equation}
W=\frac{1}{N}\Tr\cP\, 
e^{i\int (A_\mu \dot x^\mu+iy^i\Phi_i)ds}\,,
\end{equation}
where $A_\mu$ is the gauge field and $\Phi_i$ are the six scalars 
(one may include also couplings to the fermi-fields, but we do not 
write them explicitly). $x^\mu(s)$ is a closed curve (or an infinite 
one, assuming appropriate boundary conditions) and $y^i(s)$ are 
arbitrary couplings to the scalars. Here we used Euclidean 
conventions, where the scalar term is multiplied by an $i$, below 
we will also work in Lorentzian signature where for a time-like 
curve this $i$ should not be included.

If the path is an infinite straight line or a circle and if it couples 
to only one of the scalars with the appropriate strength, say 
$y^i=|\dot x|\delta^{i6}$, this 
operator will preserve half the supersymmetries of the vacuum 
\cite{Mikhailov:2002ya,Bianchi:2002gz}. While there are a lot 
of interesting results for the simple circular loop 
\cite{Erickson:2000af,Drukker:2000rr,Drukker:2005kx}, 
we wish to consider more general operators but still limit ourselves 
to operators close to the symmetric ones.

Consider a Wilson loop whose path is close to a circle of radius 
$R$ in the $(1,2)$ plane and the scalar couplings close to $\Phi_6$. 
We may write it as a deformation of the circular path as 
\begin{equation}
x^\mu(s)=x_0^\mu(s)+\epsilon^\mu(s)\,,\qquad
x_0^\mu(s)=(R\cos s,\,R\sin s,\,0,\,0)\,,\qquad
y^i(s)=|\dot x_0|\delta^{i6}+\epsilon^i(s)\,,
\end{equation}
and then expand in powers of $\epsilon(s)$. By this procedure we 
may express an arbitrary Wilson loop close to the circle as a sum 
over deformations of the basic circular loop (see for example 
\cite{Polyakov-Rychkov,Semenoff:2004qr,Miwa:2005qz})
\begin{equation}
\begin{aligned}
W[x^\mu,\,y^i]=&\ 
\Bigg(1+\int ds\left[
\epsilon^\mu(s)\frac{\delta}{\delta x^\mu(s)}
+\epsilon^i(s)\frac{\delta}{\delta y^i(s)}\right]
\\&\ \quad
+\frac{1}{2}\int ds_1\,ds_2\left[
\epsilon^\mu(s_1)\epsilon^\nu(s_2)
\frac{\delta^2}{\delta x^\mu(s_1)\delta x^\nu(x_2)}
+\cdots\right]
+O(\epsilon^3)\Bigg)
W_{\hbox{\scriptsize circle}}\,.
\end{aligned}
\end{equation}

Those deformations may, in turn, be written as local insertions into 
the loop, the functional derivatives with respect to $x^\mu(s)$ introduce 
a field-strength $F_{\mu\nu}\dot x^\nu$ or a covariant derivative 
$D_\mu$. the derivatives with respect to $y^i(s)$ insert the scalar 
field $\Phi_i$. So the small deformation of the circle may be written as
\begin{equation}
\begin{aligned}
W[x^\mu,\,y^i]=&\ 
\frac{1}{N}\Tr\cP\Bigg[\bigg(1+\int ds\Big[
i\epsilon^\mu(s)\dot x_0^\nu(s) F_{\mu\nu}(x_0(s))
-\epsilon^\mu(s)|\dot x_0|D_\mu\Phi_6(x_0(s))
\\&\ \hskip1.3in
-\epsilon^i(s)|\dot x_0|\Phi_i(x_0(s))\Big]
+O(\epsilon^2)\bigg)
e^{i\int (A_\mu \dot x_0^\mu+i|\dot x_0|\Phi_6)ds}\Bigg]\,,
\end{aligned}
\end{equation}
We did not write explicitly the $O(\epsilon^2)$ term, it is 
straight-forward to derive it, but the resulting expression is quite long. 
The only subtlety is that there is contact term, when two functional 
derivatives act at the same point in addition to two $F$s there will be 
an extra $DF$ term.

From this discussion we see that instead of considering a general path 
and general scalar couplings we can take Wilson loops with $p$ 
insertions of local operators in the adjoint representations of the gauge 
group at different positions $x_1,\cdots, x_p$ along the loop
\begin{equation}
W[\cO_p(x_p)\cdots\cO_1(x_1)]
=\frac{1}{N}\Tr\cP \left[\cO_p(x_p)\cdots\cO_1(x_1)\, 
e^{i\int (A_\mu \dot x_0^\mu+i|\dot x_0|\Phi_6)ds}\right]\,.
\end{equation}
The calculation of the expectation value of those operators may also 
be regarded as the $p$-point function of the 
adjoint operators $\cO_i$ along the circle. This interpretation relies 
on the fact that the basic circular Wilson loop is a very natural object, 
it preserves half the supersymmetries and is the most obvious way 
to connect non-singlet operators to form a gauge invariant observable. 
In this paper we will mainly concentrate on the case of two insertions, 
but also discuss a single one. 

In \cite{Drukker:2005af} we studied the symmetry of the circle and 
line in a conformal field theory and the representations of this symmetry 
group. Let us review it now.

Starting with the straight line, the subgroup of the conformal group 
$SO(5,1)$ of four dimensional Euclidean space that keeps an infinite 
straight line invariant is $SL(2,\bR)\times SO(3)$. The $SO(3)$ is 
given by rotations around the line while the generators of $SL(2,\bR)$ 
are time translation $P_t$, dilation $D$ and a special conformal 
transformation $K_t$. Those act on scalar operators by
\begin{equation}
\begin{aligned}[]
[J_+,\,\cO]=&\ [-iP_t,\,\cO]=-\partial_t\cO\,,\\
[J_0,\,\cO]=&\ [iD,\,\cO]=(\Delta+x^\mu\partial_\mu)\cO\,,\\
[J_-,\,\cO]=&\ [iK_t,\,\cO]=
(x^2\partial_t-2t x^\mu\partial_\mu-2t\Delta)\cO\,,\\
\end{aligned}
\end{equation}

The Wilson loop in $\cN=4$ gauge theory with the appropriate coupling 
to the scalar field $\Phi_6$ preserves half the supersymmetries of the 
vacuum. The even part of the full group includes in 
addition to the conformal group also the $SO(6)$ R-symmetry. The 
Wilson loop breaks it to a supergroup whose bosonic part is 
$SL(2,\bR)\times SO(3)\times SO(5)$
\cite{Bianchi:2002gz,Drukker:2005kx}. The $SL(2,\bR)$ part is 
the one written above and it is left invariant by the Wilson loop with 
no insertions (this was also noticed in \cite{kapustin}).

We may now look at other Wilson loops and ask how they 
transform under $SL(2,\bR)$. In \cite{Drukker:2005af} we studied 
this problem at tree-level and showed that for a single insertion of 
conformal dimension $\Delta$ the Wilson loop will be in a 
representation of $SL(2,\bR)$ with quadratic Casimir 
$-\Delta(\Delta-1)$. In Section~\ref{sec-local} we will take the 
first steps to include quantum corrections.

We will start, though, by considering two insertions. In that case it 
turn out to be more useful to consider the representation of each of those 
operators under $SL(2,\bR)$ rather than the full object---Wilson loop 
with two insertions. Since the Wilson loop itself does not break 
$SL(2,\bR)$, we may ask in what way each of the local insertions 
breaks it and assign to them a conformal dimension.

Consider the Wilson loop with two insertions, one at the origin and 
the other at $t$. For $t\neq0$ the Ward identity associated with the 
dilatational symmetry is
\begin{equation}
J_0\vev{W[\cO'(t)\,\cO(0)]}
=(\Delta_\cO+\Delta_{\cO'}+t\,\partial_t)\vev{W[\cO'(t)\,\cO(0)]}
=0\,.
\end{equation}
The solution to this differential equation is
\begin{equation}
\vev{W[\cO'(t)\,\cO(0)]}
\propto\frac{1}{t^{\Delta_\cO+\Delta_{\cO'}}}\,.
\label{line-2-pt}
\end{equation}

We are restricting ourselves to consider operators only along the line. We 
may still use conformal symmetry to find the form of the two-point 
function, and the main advantage is that since we have the Wilson loop 
running along that line, the operators need not be singlets of the gauge 
group but rather may transform in the adjoint representation. 
We use this method to define the dimension of an adjoint operator to be 
equal to $\Delta$ as calculated in (\ref{line-2-pt}).

The same can be done for the circle, which is related to the straight line 
by a conformal transformation. The advantage over the line is that it is 
better defined---the line is invariant only under gauge transformations that 
vanish at infinity. The generators of $SL(2,\bR)$ are now
\begin{equation}
\begin{aligned}[]
J_0=&\ -\frac{i}{2}\left(RP_1+\frac{K_1}{R}\right)\,,\\
J_\pm=&\ -iM_{12}\mp\frac{i}{2}\left(RP_2+\frac{K_2}{R}\right)\,.
\end{aligned}
\end{equation}
$P_i$ and $K_i$ are the generators of translations and conformal 
transformations in the plane of the circle and $M_{12}$ generates 
rotations in the plane.

If $\eta$ is a radial coordinate in the plane of the circle and $\zeta$ 
in the orthogonal plane, we define
\begin{equation}
\sin\theta=\frac{\zeta}{\tilde r}\,,\qquad
\sinh\rho=\frac{\eta}{\tilde r}\,,\qquad
\tilde r=\frac{\sqrt{(\zeta^2+\eta^2-R^2)^2+4R^2\zeta^2}}{2R}
=\frac{R}{\cosh\rho-\cos\theta}\,.
\label{coordinates}
\end{equation}
In terms of those coordinates the action of the $SL(2,\bR)$ generators 
on scalar operators is
\begin{equation}
\begin{aligned}[]
[J_0,\,\cO(\theta,\rho,\psi)]=&\ 
\tilde r^{-\Delta}(-\cos\psi\,\partial_\rho
+\coth\rho\sin\psi\,\partial_\psi)\tilde r^\Delta
\cO(\theta,\rho,\psi)\,,\\
[J_\pm,\,\cO(\theta,\rho,\psi)]=&\ \mp\tilde r^{-\Delta}\left[
\sin\psi\,\partial_\rho+(\cos\psi\coth\rho\pm1)\partial_\psi
\right]\tilde r^\Delta\cO(\theta,\rho,\psi)\,.
\end{aligned}
\end{equation}

For operators along the circle of radius $R$, we take $\rho\to\infty$ 
so $\tilde r\to 2Re^{-\rho}$ and the action of $J_0$ reduces to
\begin{equation}
[J_0,\,\cO(\psi)]
=(\Delta\cos\psi+\sin\psi\,\partial_\psi)\cO(\psi)\,,
\end{equation}
hence the Ward identity for a loop with two insertions at $0$ and 
$\psi$ is
\begin{equation}
J_0\vev{W[\cO'(\psi)\,\cO(0)]}
=(\Delta_\cO+\Delta_{\cO'}\cos\psi+\sin\psi\partial_\psi)
\vev{W[\cO'(\psi)\,\cO(0)]}
=0\,.
\end{equation}
This is solved by\footnote%
{We thank Harald Dorn for pointing out an error in the original 
version of this manuscript.}
\begin{equation}
\vev{W[\cO'(\psi)\,\cO(0)]}
\propto\frac{\cos^{|\Delta_\cO-\Delta_{\cO'}|}(\psi/2)}
{\sin^{\Delta_\cO+\Delta_{\cO'}}(\psi/2)
}\,.
\end{equation}
In the case where the two operators have the same dimension the 
denominator is a power of $\sin(\psi/2)$, and after rescaling 
the proportionality constant by $(2R)^{-2\Delta}$, we get that the 
expectation value is just a power of the distance between the two 
insertions at $0$ and at $\psi$. 

Note that in \cite{Drukker:2005af} we focused mainly on another 
basis, where $J_0= i\partial_\psi$. In that basis the eigenstates 
of $J_0$ are Wilson loops with Fourier modes of those insertions, 
i.e. they are smeared around the circle with phase factors 
$e^{im\psi}$. That basis is natural for some purposes, particularly 
when studying the representation of a loop with a single insertion, but 
it also has some difficulties, for example the representations we find 
in that basis are generally non-unitary. In this paper we will use 
the basis written above%
\footnote{N.D. would like to thank Zack Guralnik for the inspiration 
to look at this basis.} 
which seems more appropriate for the study 
of local insertions into the loop. Insertions at $\psi=0$ form 
highest-weight representations of $SL(2,\bR)$.

\section{Gauge theory calculation}

\subsection{Tree-level}

Let us start performing explicit calculations. We consider the two 
complex combinations of the scalar fields
\begin{equation}
Z=\frac{1}{\sqrt2}(\Phi_1+i\Phi_2)\,,
\qquad
X=\frac{1}{\sqrt2}(\Phi_3+i\Phi_4)\,.
\end{equation}
Note that we chose them so they will not contract at tree-level with 
$\Phi_6$ which appears in the phase factor of the Wilson loop.

We will insert two operators transforming in the adjoint representation 
into the Wilson loop. The equations below are written for the case of 
the straight line, but they are essentially the same for the circle. 
We take one operator $\cO$ at the origin 
to be a word made of  the letters $Z$ and $X$ and another operator, 
$\cO^{\prime\dagger}$, made of the 
complex conjugates $\bar Z$ and $\bar X$ inserted at $t$.
Explicitly we have
\begin{equation}
W[\cO^{\prime\dagger}(t)\,\cO(0)]=\frac{1}{N}
\Tr\cP\left[\cO^{\prime\dagger}(t)\,\cO(0)\,
e^{i\int (A_t+i\Phi_6)dt}\right]\,.
\label{two-insertions}
\end{equation}

Taking the standard scalar propagator, which is proportional to the 
identity matrix on the color indices amounts to a gauge choice which 
turns out to be very convenient for our calculation; at tree-level the 
holonomy will not contribute. Thus at leading order the expectation 
value of the Wilson loop will involve just the contraction of those 
two words. In effect it is
\begin{equation}
\vev{\frac{1}{N}\Tr\left[
\cO^{\prime\dagger}(t)\,\cO(0)\right]}\,.
\end{equation}

There is a single trace over the two words at the origin and at $t$, so 
the only planar contribution at tree-level is the contraction of the first 
letter of the operator at the origin with the last of the operator at $t$, 
the second at the 
origin with the next-to-last at $t$ and so on. Each contraction comes 
with a Kronecker delta on the flavor index, $Z$ or $X$ and factor 
of $\lambda/8\pi^2t^2$, where $\lambda=g_{YM}^2N$ is the 't Hooft 
coupling. So the final answer at tree-level may be written in matrix 
notations as
\begin{equation}
\vev{W[\cO^{\prime\dagger}(t)\,\cO(0)]}\propto
\left(\frac{\lambda}{8\pi^2t^2}\right)^K
I\,,
\label{tree}
\end{equation}
where $K$ is the length of the word and $I$ is the identity matrix on 
the flavor indices. The relevant graph is shown in figure~\ref{tree-fig}.

\figin{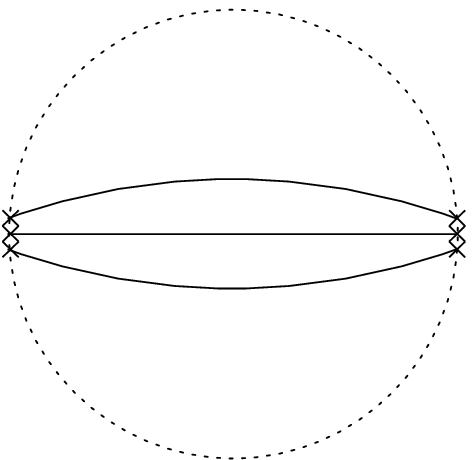}{The tree-level diagram for a circular Wilson loop 
(the dotted line) with the insertion of two words each made of three 
scalars. The Wilson loop sets the order at which the gauge indices are 
contracted and only the depicted diagram is planar.}

If we were studying the 2-point function of single-trace local operators 
we would have to account for the cyclicity of the trace, so $\Tr ZZX$ 
for example, would have a nonzero contraction with 
$\Tr\bar Z\bar X\bar Z$ at tree-level even in the planar approximation. 
In our case they would not, and since local operators are described in 
terms of periodic spin-chains, it seems like the insertions into the Wilson 
loop will be described by open spin-chains.

A local operator which is the trace of a word of length $K$ has dimension 
$\Delta=K$ classically. Similarly the insertion of the word of length 
$K$ at the origin has classical dimension (i.e. eigenvalue of $iD$ or $J_0$) 
$\Delta=K$ as can be seen from the exponent of $t$ in the classical 2-point 
function (\ref{tree}).

\subsection{One-loop}
\label{sec-one-loop}
Beyond tree-level the Feynman diagrams will generically diverge 
and we will have to renormalize the Wilson loops. The theory of 
renormalization of Wilson loop operators is quite complicated 
\cite{Brandt:1981kf}. In general there will be a linear divergence 
proportional to the circumference of the loop and in addition, if the 
curve is not smooth there may be logarithmic divergences. Such 
divergences arise also when the operator has end-points (with 
quark insertions) \cite{Craigie:1980qs,Aoyama:1981ev,Knauss:1984rx} 
and the same happens from the insertion of adjoint operators (see 
\cite{Dorn:1986dt}).

In our case the inclusion of the coupling to the scalar $\Phi_6$ in 
the exponent guarantees that without the local insertions the Wilson 
loop is a finite operator that does not require renormalization. The 
local insertions change that, they lead to logarithmic divergences 
in perturbation theory. The divergences associated with cusps and 
intersections in the loop are well studied and depend on the angle 
of the cusp. In a similar way the divergences coming from the local 
insertion will depend on the insertion. We have not explored all 
possible divergences coming from such insertions at high orders 
in perturbation theory. We contend ourselves for now with analyzing 
only the relevant one-loop diagrams.

Our prescription for the renormalization of the Wilson loop 
will involve multiplicative renormalization of each of the insertions
\begin{equation}
W_{\rm ren}[\cO_p\cdots\cO_1]
=Z_{\cO_p}\cdots Z_{\cO_1}W[\cO_p\cdots\cO_1]\,.
\end{equation}
In the usual fashion those $Z_{\cO_i}$ factors will cancel the divergences 
that come from subgraphs that approach the insertion $\cO_i$ rendering 
$W_{\rm ren}$ finite.

In a conformal field theory we associate divergences with the 
renormalization of the conformal dimension. We propose the same 
interpretation here, and since the renormalization factors are 
associated with each individual insertion, this means that we should 
associate a conformal dimension to each insertion. This is the 
justification for this interpretation that was presented in the preceding 
sections. If there is a single insertion into the loop we are free to 
associate the conformal dimension either with the insertion or with 
the entire Wilson loop.

Going back to the line with two insertions, let us consider as a first 
example each of the insertions to be just a single scalar field, the first 
$Z(0)$ and the other $\bar Z(t)$. There are two graphs that contribute 
at the one-loop level; the self-energy graph and a graph where the scalar 
propagator exchanges a gluon with the Wilson loop (like in 
figure~\ref{boundary-fig}). Each of these 
graphs diverges, but together the divergences exactly cancel as can 
be extracted from the calculations in \cite{Erickson:2000af}. We 
review this calculation in Appendix~\ref{sec-calc}.

From the argument above we know that a Wilson loop with the 
insertion of any number of scalar fields $Z$ or $X$ (and their 
conjugates) will have a finite expectation value as long as none of 
the scalar insertions are at the same point. For coincident scalars there 
will be more divergences coming from the exchange of gluon between 
the two scalar propagators (H-graph) and from the quartic scalar 
interaction vertex (X-graph). Those divergences will lead to a 
non-trivial renormalization of the insertions that are made up of 
more than one scalar.

All those graphs include only the interactions between the scalars 
in the insertion and do not involve the Wilson loop (see 
figure~\ref{bulk-fig}). So they were 
all evaluated before in the context of calculating the one-loop 
corrections to the dimensions of local operators. See for example 
\cite{Berenstein:2002jq,Kristjansen:2002bb,Constable:2002hw,
Gross:2002mh}.

\figin{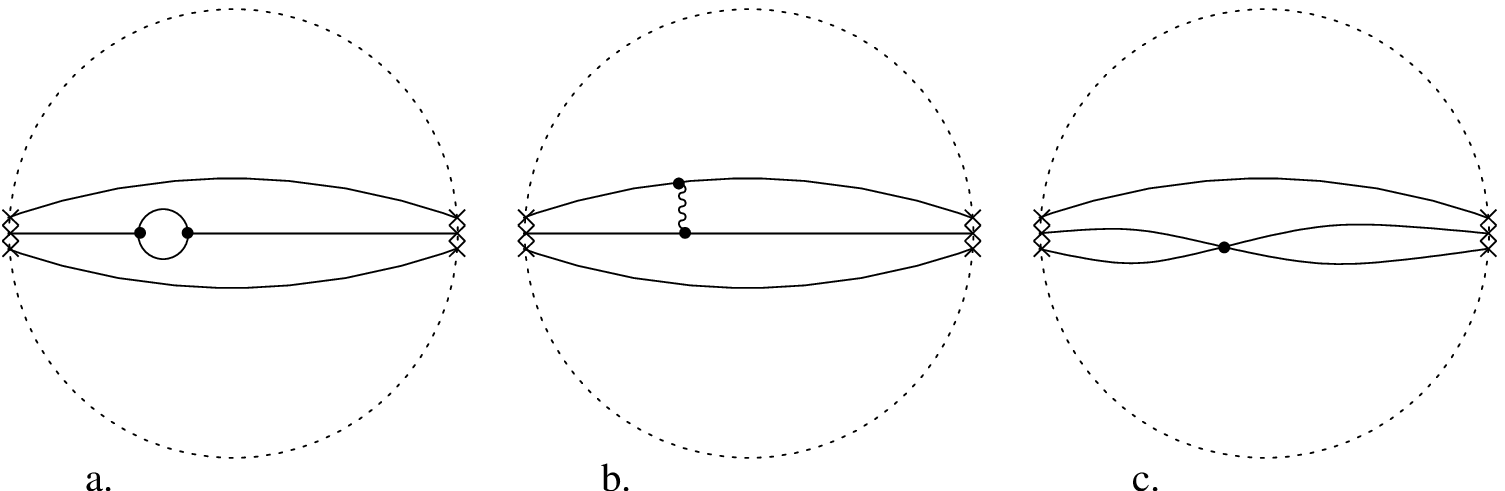}{The planar one-loop graphs that do not involve the 
Wilson loop are the same as for single trace local operators. The 
self energy diagrams like (a) includes all possible fields going around 
the loop. The H-diagrams, like (b), involves the exchange of a gluon 
between nearest-neighbors. The X-diagrams, like (c), involves the 
quartic scalar term and leads to the permutation term in the spin-chain 
Hamiltonian. Note though, that the H and X interaction graphs 
involving the first and last scalars are not planar and therefore are 
not included.}

The $Z$-factor associated with the H-graph is
\begin{equation}
Z_{\rm H}=I-\frac{\lambda}{16\pi^2}\ln\Lambda\,I\,,
\end{equation}
where $I$ is the identity in flavor-space and $\Lambda$ a UV-cutoff. 
If the insertion is made of $K$ scalar fields there will be $K-1$ planar 
H-graphs, connecting nearest-neighbors.

The X-diagrams mix between nearest-neighbors and their divergences 
are compensated by the $Z$-factor
\begin{equation}
Z_{\rm X}=I+\frac{\lambda}{16\pi^2}(I-2P)\ln\Lambda\,,
\end{equation}
where $P$ permutes the two scalars. Again there will be $K-1$ such 
graphs.

Next we have the self-energy graphs. we associate half of those 
divergences with the composite insertion. The other half will be 
associated with the other end of the propagator (and may be 
canceled by the divergences in the interaction with the Wilson loop, 
as explained above). There are $K$ such graphs, each giving a 
renormalization factor
\begin{equation}
Z_{\hbox{\scriptsize self-energy}}
=I+\frac{\lambda}{8\pi^2}\ln\Lambda\,I\,.
\end{equation}

Then there are the graphs involving the Wilson loop and in the 
planar limit will connect only to the outermost scalars in the 
insertion with the loop (see figure~\ref{boundary-fig}). 
Those are exactly the same graphs that 
appeared when there were isolated scalars along the loop. So those 
graphs contribute a $Z$-factor that cancels a single self-energy one
\begin{equation}
Z_{\hbox{\scriptsize boundary}}
=I-\frac{\lambda}{8\pi^2}\ln\Lambda\,I\,.
\label{z-boundary}
\end{equation}
Note that those graphs too are indifferent to the flavor, i.e. they 
are the same for an $X$ or a $Z$ insertion.

\figin{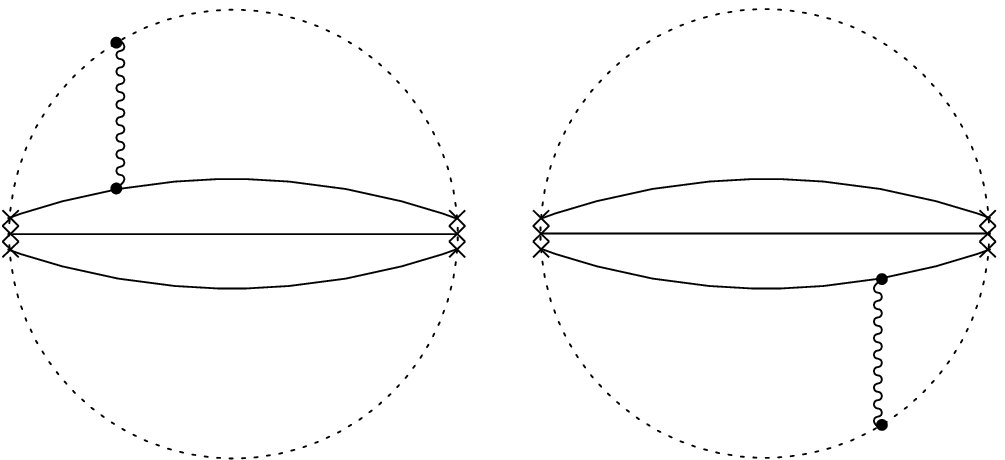}{At the planar one-loop level only the external most 
lines may interact with the Wilson loop. These diagrams come from 
expanding the holonomy to first order bringing down the gauge field 
and $\Phi_6$. This gauge field can then be contracted with the 
outermost scalars. Note that the scalars~$Z$ and~$X$ do not contract 
with~$\Phi_6$.}

Combining all those we find the total one-loop renormalization factor
\begin{equation}
Z_{\hbox{\scriptsize total}}
=I+\frac{\lambda}{8\pi^2}\ln\Lambda
\sum_{k=1}^{K-1}(I-P_{k,k+1})\,.
\end{equation}

\subsection{Spin-chain interpretation and the Bethe ansatz}
\label{sec-Bethe}

The matrix of anomalous dimensions is given by
\begin{equation}
\Gamma=\frac{1}{Z}\frac{dZ}{d\ln\Lambda}
\sim\frac{\lambda}{8\pi^2}\sum_{k=1}^{K-1}(I-P_{k,k+1})\,.
\end{equation}
Eigenvectors of this mixing matrix will undergo only multiplicative 
renormalization which is due to the anomalous dimension. From 
the Ward identity for insertions into the line the correlator of two 
insertions $\cO_n$ of length $K$ with eigenvalues $\gamma_n$ of 
$\Gamma$ is
\begin{equation}
\vev{W[\cO_n^\dagger(t)\,\cO_n(0)]}\sim
\frac{1}{t^{2(K+\gamma_n)}}\,.
\end{equation}
Here $\cO_n^\dagger$ is the operator with the complex conjugate 
fields in the reverse order.

So at one loop the anomalous dimensions of the insertions into the loop 
are given by the eigenvalues of $\Gamma$. As presented in 
\cite{Minahan:2002ve}, this matrix may be regarded as the Hamiltonian 
of a one-dimensional spin-chain. In their case it was a periodic chain, 
but in ours it's open, it starts at $k=1$ and ends at $k=K$. The total 
number of $Z$ and $X$ insertions is fixed as they cannot move off 
the chain, so we have purely reflective, or Neumann boundary conditions.

It turn out this system is integrable and it is well known how to 
diagonalize its Hamiltonian by use of the Bethe ansatz. Actually very 
similar open spin-chains were found to describe the anomalous 
dimension of operators in a variety of different systems 
\cite{Stefanski:2003qr,Chen:2004mu,DeWolfe:2004zt,Chen:2004yf,
Susaki:2004tg,Berenstein:2005vf,Berenstein:2005fa,McLoughlin:2005gj,
Susaki:2005qn,Erler:2005nr,Beisert:2005he,Okamura:2005cj,
Agarwal:2006gc,Okamura:2006zr}. 
Most of those papers consider systems with fundamental fields, either 
by taking a gauge theory with $\cN=2$ supersymmetry dual 
to an orbifold of the pp-wave geometry or by looking at defect 
CFTs, dual to string theory with a D-brane inside $AdS_5\times S^5$. 
The fundamental fields at the beginning and end of the word are 
a simple way to construct a gauge invariant operator which is 
not periodic.

Our system does not require adding extra degrees of freedom to the 
theory, since any gauge theory contains Wilson loop observables. 
In that regard it is similar to the construction of 
\cite{Berenstein:2005vf,Berenstein:2005fa} which relied on 
the determinant operator dual to a maximal giant graviton. That 
observable too is an object within $\cN=4$ gauge theory and not 
a deformation of it.

Though we can quote the results of the other papers that considered 
open spin-chains, let us be pedagogical and go over the steps of solving 
the open spin-chain.

One obvious eigenstate of the Hamiltonian is the state with all $Z$s (or 
all $X$s), this is the ferromagnetic vacuum. This state has vanishing 
anomalous dimension and is supersymmetric.

Then we can look at states with a single $X$ among all the $Z$s. 
The Hamiltonian will shift its position along the chain, so we expect 
the solutions to be standing waves. Take the superposition 
of states with the $X$ inserted at position $k$
\begin{equation}
\ket{\psi}=\sum_{k=1}^K\cos p(k-1/2)\,\ket{k}\,,
\end{equation}
with $p=n\pi/K$ for integer $n<K$ (this should be thought of as the 
lattice version of Neumann boundary conditions). These are all 
eigenstates of the Hamiltonian where the only subtlety comes from 
treating the boundaries
\begin{equation}
\begin{aligned}
\Gamma\ket{\psi}=&\ \frac{\lambda}{8\pi^2}\Bigg[
\sum_{k=2}^{K-2}
\Big[2\cos p(k-1/2)-\cos p(k-3/2)-\cos p(k+1/2)\Big]\ket{k}
\\&\ \qquad
+\Big[\cos(p/2)-\cos(3p/2)\Big]\ket{1}
+\Big[\cos p(K-1/2)-\cos p(K-3/2)\Big]\ket{K}\Bigg]
\\=&\ 
\frac{\lambda}{2\pi^2}\sin^2\frac{p}{2}\,\ket{\psi}\,.
\end{aligned}
\end{equation}
The anomalous dimensions 
of those operators with a single $X$ insertion are thus
\begin{equation}
\gamma_n=\frac{\lambda}{2\pi^2}\sin^2\frac{n\pi}{2K}
\sim\frac{\lambda n^2}{8K^2}\,.
\end{equation}
The last expression is valid for $K\gg n$.

The story gets more complicated when considering more $X$ insertions, 
so we use the methods of the algebraic Bethe ansatz 
\cite{Bethe:1931hc} (see for example \cite{Faddeev:1996iy}). 
The theory of open spin-chains is well developed but for our purpose 
it will not be necessary to use those techniques. Since our boundary 
conditions are purely reflective we can simply add an image chain with 
$k=K+1,\cdots 2K$, and consider the periodic chain of length $2K$.

Since the construction has to be symmetric under reflections, for every 
impurity at position $k$ we have to place another one at position 
$2K+1-k$, or in the momentum basis require symmetry under 
$p\to -p$. The Hamiltonian of the regular closed Heisenberg chain of 
length $2K$ will include in addition to the interaction terms between the 
spins at positions $k=1,\cdots,K$ and their images at $K+1,\cdots,2K$
also the terms acting on positions $K$, $K+1$ and $2K$, $1$
\begin{equation}
\frac{\lambda}{8\pi^2}
\left(I-P_{K,K+1}+I-P_{2K,1}\right)\,.
\end{equation}
Due to the reflection symmetry the spin at position $K$ and $K+1$ 
are always equal and the same is true for the other pair. So this extra term 
vanishes and we may use the regular Hamiltonian for the periodic spin-chain 
of length $2K$ and by imposing reflection symmetry we will find the 
spectrum of the open spin-chain.

The Bethe equation for a closed chain is given in terms of 
the Bethe roots related to the momenta by
\begin{equation}
u_k=\frac{1}{2}\cot\frac{p_k}{2}\,.
\end{equation}
For a chain of length $2K$ and $2M$ impurities the equations are
\begin{equation}
\left(\frac{u_j+i/2}{u_j-i/2}\right)^{2K}
=\prod_{\genfrac{}{}{0pt}{}{k=1}{k\neq j}}^{2M}
\frac{u_j-u_k+i}{u_j-u_k-i}\,.
\end{equation}
The right hand side corresponds to the interaction of the impurity 
$j$ with all the other impurities. In the last equation we considered 
an arbitrary distribution of impurities, but reflection symmetry 
requires that those impurities form pairs with opposite momenta, 
so $u_{M+1}=-u_1$ and so on. Accounting for that, the last equation 
reads
\begin{equation}
\left(\frac{u_j+i/2}{u_j-i/2}\right)^{2K}
=\prod_{\genfrac{}{}{0pt}{}{k=1}{k\neq j}}^M
\frac{(u_j-u_k+i)(u_j+u_k+i)}{(u_j-u_k-i)(u_j+u_k-i)}\,.
\label{Bethe-equation}
\end{equation}
For consistency, for any solution to this system with a positive $u_j$ 
there has to be another one with negative root, and this indeed holds.

The anomalous dimension for any solution is given by the sum over 
the individual impurities (counting every pair of impurities only once!)
\begin{equation}
\gamma_n=\frac{\lambda}{2\pi^2}\sum_{k=1}^M\sin^2\frac{p_k}{2}
=\frac{\lambda}{8\pi^2}\sum_{k=1}^M\frac{1}{u_k^2+1/4}\,.
\end{equation}

With a single impurity, or $M=1$ we find the equation
\begin{equation}
\left(\frac{u+i/2}{u-i/2}\right)^{2K}=1
\qquad\Rightarrow\qquad
u=\frac{1}{2}\cot\frac{\pi n}{2K}\,.
\end{equation}
Those are the same values of the momenta we found before and as 
stated the anomalous dimensions are
\begin{equation}
\gamma_n=\frac{\lambda}{2\pi^2}\sin^2\frac{\pi n}{2K}
\sim\frac{\lambda n^2}{8K^2}\,.
\label{bethe-spectrum}
\end{equation}
Later we will compare this to the $AdS$ result.

\subsection{The thermodynamic limit}
\label{sec-thermo}

We can also solve the Bethe equations in the thermodynamic limit, 
when $K\to\infty$ and $M\to\infty$ with a fixed ratio $M/K$. 
In that limit the roots condense into cuts in the complex plane. This 
was studied in 
\cite{Beisert:2003xu,Beisert:2003ea,Arutyunov:2003rg,
Engquist:2003rn} 
for closed spin-chains, dual to closed strings. In those examples they 
always took solutions that were symmetric under $u\to-u$, so for 
every cut on the right of the imaginary axis there was a mirror cut 
on the left.

This was done to guarantee that the total momentum vanishes, though 
this is a much stronger constraint. In our case this is exactly the 
symmetry required so those solutions of the closed chain are also 
solutions of our open spin-chains. We can copy their results 
remembering to take only one of the cuts into account.

To review the solution, consider the logarithm of 
(\ref{Bethe-equation})
\begin{equation}
2K\ln\frac{u_j+i/2}{u_j-i/2}
=2\pi i n_j+\sum_{\genfrac{}{}{0pt}{}{k=1}{k\neq j}}^M
\ln\frac{(u_j-u_k+i)(u_j+u_k+i)}{(u_j-u_k-i)(u_j+u_k-i)}\,.
\end{equation}
$n_j$ are arbitrary integers, corresponding to different branches 
of the logarithm. We will focus on the solution where all the $n_j$ 
are equal (the image roots of course have $-n_j$). In the large 
$K$ limit we may look at solutions were all $u_j$ 
scale with $K$. In that limit, after rescaling back to finite quantities 
the equation becomes
\begin{equation}
\frac{1}{u_j}
=\pi\,n_j+\frac{2}{K}
\sum_{\genfrac{}{}{0pt}{}{k=1}{k\neq j}}^M
\frac{u_j}{u_j^2-u_k^2}\,.
\end{equation}

In this limit the roots form smooth curves on the complex plane around 
$\pi\,n_j$, where we may introduce the root density
\begin{equation}
\rho(u)=\frac{1}{M}\sum_{j=1}^M\delta(u-u_j)\,.
\end{equation}
We label $\cC$ the contour along which the eigenvalues are distributed. 
By the definition $\rho(u)$ is normalized to
\begin{equation}
\int_\cC du\,\rho(u)=1\,,
\end{equation}
and the equations for the roots are represented by the singular integral 
equation (with principle part prescription)
\begin{equation}
\frac{2M}{K}-\hskip-13pt\int_\cC
dv\,\frac{\rho(v)u^2}{u^2-v^2}
=1-\pi nu\,.
\end{equation}
The anomalous dimension is
\begin{equation}
\gamma_M=\frac{\lambda}{8\pi^2}\frac{M}{K^2}
\int_\cC du\,\frac{\rho(u)}{u^2}\,.
\end{equation}

The solution to  this equation involves analytically continuing to 
negative $M$ where the eigenvalues are all real, so the contour 
$\cC$ is given by an interval on the real axis. For the details of the 
solution we refer the reader to \cite{Beisert:2003xu,Beisert:2003ea}. 
We just quote the final answer (for $n_j=1$), that the anomalous 
dimension is given by the solution to the transcendental equations 
involving elliptic integrals of the first and second kind%
\footnote{The only difference from 
\cite{Beisert:2003xu,Beisert:2003ea} are a few factors of 2, 
because we consider only the roots to the right of the imaginary 
axis. Also they use a definition of the elliptic integral whose modulus 
is the square of our $k$.}
\begin{equation}
\begin{aligned}
\gamma_M=&\ \frac{\lambda}{8\pi^2K}K(k)\Big(
2E(k)-(2-k^2)K(k)\Big)\,,\\
\frac{M}{K}=&\ 
\frac{1}{2}-\frac{1}{2\sqrt{1-k^2}}\frac{E(k)}{K(k)}\,.
\end{aligned}
\label{elliptic}
\end{equation}

\section{String-theory description}
\label{sec-strings}

Let us now describe the same system, a Wilson loop with two insertions 
in the dual string theory on $AdS_5\times S^5$. As in the BMN 
construction \cite{Berenstein:2002jq} we map one of the insertions 
to the infinite past in global Lorentzian $AdS_5$ and the other one 
to future infinity. Those local insertions are connected by the Wilson 
loop which in this picture will run up and down two lines at 
antipodal points on the $\bR\times S^3$ boundary.

To describe the Wilson loop in supergravity we have to find a solution 
of the classical string equations of motion that satisfies the following 
properties:
\begin{enumerate}
\item
It should extend to the two lines on the boundary of $AdS_5$ and 
when it reaches the boundary it should be at the point of $S^5$ that 
corresponds to the scalar $\Phi_6$ (we'll call that the north pole).
\item
Depending on the number of $Z$s and $X$s the world sheet should 
carry angular momentum around two orthogonal angular directions 
on $S^5$, $\phi_1$ and $\phi_2$ (which are also orthogonal to the 
north pole).
\item
For the operator with a large number of $Z$s and no $X$s, as the 
surface gets close to the center of $AdS_5$, it should approach the 
large circle in the $\phi_1$ direction on $S^5$ 
(the equator). Then one can zoom in on part of the world-sheet near 
the center of $AdS$ and the equator, and take a Penrose limit. 
In that limit it should reduce to a solution of string theory on the maximally 
supersymmetric pp-wave background. Some of the excitations 
of that solution would correspond to $X$ impurities.
\end{enumerate}

The metric of $AdS_5$ of curvature radius $L$ in global coordinates is
\begin{equation}
ds^2
=L^2\left[-\cosh^2\rho\,dt^2+d\rho^2
+\sinh^2\rho\,d\Omega_3^2\right]\,,
\end{equation}
where $d\Omega_3^2$ is the metric on a unit size $S^3$. On the $S^5$ 
we will restrict our ansatz to an $S^4$ subspace with metric
\begin{equation}
ds^2
=L^2\left[d\theta^2+\sin^2\theta\left(d\psi^2
+\cos^2\psi\,d\phi_1^2+\sin^2\psi\,d\phi_2^2\right)\right]\,,
\end{equation}

The relevant part of the Green-Schwarz string action is
\begin{equation}
\begin{aligned}
\cS=&\ \frac{L^2}{4\pi\alpha'}\int d\sigma\,d\tau\,
\sqrt{-h}h^{\alpha\beta}\Big[
-\cosh^2\rho\,\partial_\alpha t\,\partial_\beta t
+\partial_\alpha\rho\,\partial_\beta\rho
+\partial_\alpha\theta\,\partial_\beta\theta
\\&\hskip1.1in
+\sin^2\theta\left(
\partial_\alpha\psi\,\partial_\beta\psi
+\cos^2\psi\,\partial_\alpha\phi_1\,\partial_\beta\phi_1
+\sin^2\psi\,\partial_\alpha\phi_2\,\partial_\beta\phi_2
\right)\Big].
\end{aligned}
\end{equation}

Now we solve the equations of motion stemming from this action like in 
\cite{Frolov:2002av,Frolov:2003qc,Tseytlin:2003ii,Drukker:2005cu}, 
by assuming a periodic ansatz
\begin{equation}
\begin{gathered}
\rho=\rho(\sigma)\,,\qquad
\theta=\theta(\sigma)\,,\qquad
\psi=\psi(\sigma)\,,\\
t=\omega\tau\,,\qquad
\phi_1=w_1\tau\,,\qquad
\phi_2=w_2\tau\,.
\end{gathered}
\end{equation}
The string Lagrangean in the conformal gauge reduces to
\begin{equation}
\cL=\frac{L^2}{4\pi\alpha'}\left[
(\rho')^2+\omega^2\cosh^2\rho
+(\theta')^2+\sin^2\theta\left(
(\psi')^2-w_1^2-(w_2^2-w_1^2)\sin^2\psi\right)\right]\,.
\end{equation}

\subsection{Solution with one angular momentum}

Let us search first for solutions carrying angular momentum in one 
direction, so we try an ansatz with $\psi=0$ and $w_2=0$. The 
Lagrangean is
\begin{equation}
\cL=\frac{L^2}{4\pi\alpha'}\left[
(\rho')^2+\omega^2\cosh^2\rho
+(\theta')^2-w_1^2\sin^2\theta\right]\,,
\end{equation}
and the equations of motion are
\begin{equation}
\begin{aligned}
\rho''-\omega^2\cosh\rho\sinh\rho=&\ 0\,,\\
\theta''+w_1^2\cos\theta\sin\theta=&\ 0\,.
\label{one-J-eom}
\end{aligned}
\end{equation}
Those may be immediately integrated to
\begin{equation}
-(\rho')^2+\omega^2\cosh^2\rho=
(\theta')^2+w_1^2\sin^2\theta=\kappa^2\,.
\end{equation}
The two integrals of motion have to be equal to each other due to the 
Virasoro constraint.

The equation for the $AdS$ coordinate $\rho$ is very simple and to 
get a single smooth surface that extends from the boundary to the center 
of $AdS_5$ and back one has to impose $\omega=\kappa$ and 
the solution is
\begin{equation}
\sinh\rho=\frac{1}{\sinh \kappa\sigma}\,.
\label{one-J-AdS-solution}
\end{equation}
For the $S^5$ coordinate, if $w_1<\kappa$ the solution will wrap the 
$S^2$ an infinite number of times while for $w_1\geq\kappa$ it will 
oscillate an infinite number of times between $\theta=0$ and the 
maximal value at $\sin\theta_0=\kappa/w_1$. 
In those cases the solution will be given by elliptic integrals, but 
the desired solution has $w_1=\kappa=\omega$ and then
\begin{equation}
\sin\theta=\tanh \kappa\sigma=\frac{1}{\cosh\rho}\,.
\label{one-J-S-solution}
\end{equation}

\figin{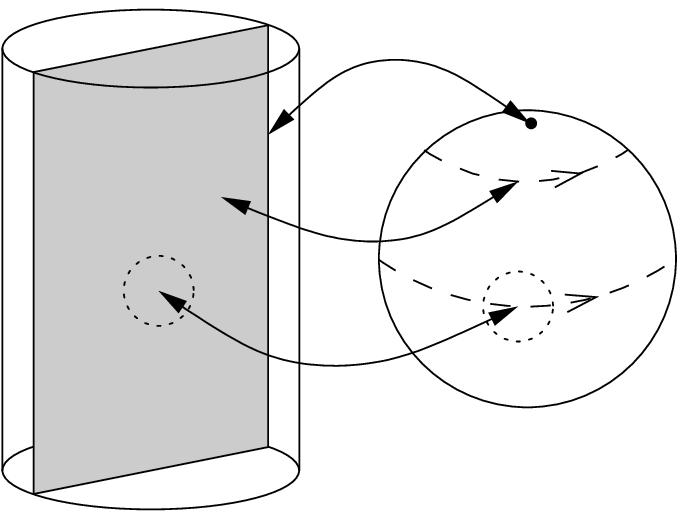}{A depiction of the string solution on $AdS_5\times S^5$. 
The string fills an $AdS_2$ subspace of $AdS_5$ (on the left). Near the 
boundary of $AdS$ it is located at the north pole of an $S^2\subset S^5$ 
(on the right). Away from the boundary it is no longer at the north pole 
and rotates around the sphere and as it gets close to the center of $AdS$ 
it approaches the equator. The dotted circles represent the region one 
zooms on to get the BMN limit.}

Let us verify that this solution satisfies the correct boundary 
conditions. At $\sigma=0$ we find $\theta=0$ and 
$\rho\to\infty$. This means that the surface approaches the 
boundary of $AdS_5$ as specified and on the sphere gets closer 
to the north-pole associated to the scalar $\Phi_6$.
As $\sigma\to\infty$ we get $\rho\sim0$ and $\theta\to\pi/2$, 
so as the string comes close to the center of $AdS_5$, it 
gets to the equator of $S^2$ and rotates around it. This range of 
$\sigma$ covers only half the world-sheet and we should analytically 
continue to negative $\sigma$ beyond $\sigma\to\infty$ to 
describe the part of the string extended in the other direction in 
$AdS_5$ (also allowing for negative $\rho$). We illustrate this 
surface in figure~\ref{ads-fig}.

The total angular momentum carried by the string is given by the 
integral (including both branches of the solution with positive and 
negative $\sigma$)
\begin{equation}
J=
\int P_\phi=
\frac{L^2}{\pi\alpha'}
\int_0^{\sigma_{\max}}d\sigma\,
\sin^2\theta\,\dot\phi
=\frac{\sqrt\lambda}{\pi}
\left(\kappa\sigma_\text{max}
-\tanh\kappa\sigma_{\max}\right)\,.
\label{one-spin-J}
\end{equation}
Here $\sigma_{\max}$ is a cutoff on the length of the world-sheet. The 
energy carried by the string is
\begin{equation}
E=
\int P_t=
\frac{L^2}{\pi\alpha'}
\int_0^{\sigma_{\max}}d\sigma\,
\cosh^2\rho\,\dot t
=\frac{\sqrt\lambda}{\pi}
\left(\kappa\sigma_\text{max}
-\coth\kappa\sigma_\text{max}
+\cosh\rho_0\right)\,.
\label{one-spin-E}
\end{equation}
$\rho_0$ is a regulator at large $\rho$ (small $\sigma$), but 
this divergence is removed by an extra boundary contribution, yielding 
in the limit of large $\kappa\sigma_\text{max}$ the result
\begin{equation}
E=J\,.
\end{equation}

\subsection{Supersymmetry}

We show now that this solution carrying angular momentum around 
a single circle preserves $1/4$ of the supersymmetries of the background. 
Using the vielbeins (only for the directions that are turned on)
\begin{equation}
e^0=L\cosh\rho\,dt\,,\qquad
e^1=L\,d\rho\,,\qquad
e^5=L\,d\theta\,,\qquad
e^6=L\sin\theta\,d\phi\,.
\end{equation}

$\Gamma_a$ will be ten real constant gamma matrices and we define 
$\gamma_\mu=e_\mu^a\Gamma_a$ and 
$\Gamma_\star=\Gamma^0\Gamma^1\Gamma^2\Gamma^3\Gamma^4$ 
the product of all the gamma matrices in the $AdS_5$ directions. With 
this the dependence of the Killing spinors on the relevant coordinates 
may be written as (see for example 
\cite{Lu:1998nu,Skenderis:2002vf,Kim:2002tj,Canoura:2005uz})
\begin{equation}
\epsilon=e^{-\frac{i}{2}\rho\,\Gamma_\star\Gamma_1}
e^{-\frac{i}{2}t\,\Gamma_\star\Gamma_0}
e^{-\frac{i}{2}\theta\,\Gamma_\star\Gamma_5}
e^{\frac{1}{2}\phi\,\Gamma_{56}}\epsilon_0\,,
\end{equation}
where $\epsilon_0$ is a chiral complex 16-component spinor. This 
satisfies the Killing spinor equation%
\footnote{
$D_\mu=\partial_\mu+\frac{1}{4}\omega_\mu^{ab}\Gamma_{ab}$
and the only relevant non-zero components of the spin-connection
are $\omega_t^{01}=\sinh\rho$ and $\omega_\phi^{56}=-\cos\theta$.}
\begin{equation}
\left(D_\mu+\frac{i}{2L}\Gamma_\star\gamma_\mu\right)
\epsilon=0\,.
\end{equation}

The projector associated with a fundamental string in type IIB is
\begin{equation}
\Gamma=\frac{1}{\sqrt{-g}}\,
\partial_\tau x^\mu\partial_\sigma x^\nu
\gamma_\mu\gamma_\nu K\,,
\end{equation}
where $g$ is the induced metric on the world-sheet and $K$ acts on 
spinors by complex conjugation. The number of supersymmetries 
preserved by the string is the number of independent solutions to the 
equation $\Gamma\epsilon=\epsilon$.

For our surface
\begin{equation}
\Gamma=\frac{\cosh^2\rho}{(\cosh^2\rho+1)\sinh\rho}
\left(-\cosh\rho\,\Gamma_{01}+\Gamma_{05}
-\sin\theta\,\Gamma_{61}
+\sin^2\theta\,\Gamma_{65}\right)K\,.
\end{equation}
The equation has to hold for all $\sigma$ and $\tau$. Since 
$\Gamma_\star\Gamma_0$ commutes with $\Gamma_\star\Gamma_5$ and 
with $\Gamma_{56}$ and also $\Gamma_\star\Gamma_1$ commutes with 
$\Gamma_\star\Gamma_5$ we may write the Killing spinor as
\begin{equation}
\epsilon=e^{-\frac{i}{2}\rho\,\Gamma_\star\Gamma_1
-\frac{i}{2}\theta\,\Gamma_\star\Gamma_5}
e^{-\frac{i}{2}\omega\tau(\Gamma_\star\Gamma_0
+i\Gamma_{56})}\epsilon_0\,.
\label{Killing-2}
\end{equation}
Since $\Gamma$ does not depend on $\tau$, the only place $\tau$ appears 
in the projector equation is in the second exponential of this expression for 
the Killing spinors. To eliminate this dependence we impose the condition
\begin{equation}
\Gamma_\star\Gamma_{056}\epsilon_0=-i\epsilon_0\,.
\label{SUSY-cond-1}
\end{equation}

Now commuting the terms in the projector $\Gamma$ through the 
remaining exponential in (\ref{Killing-2}), remembering that 
$K$ acts by complex conjugation, we get
\begin{equation}
\begin{aligned}
\Gamma\epsilon=&\ \frac{\cosh^2\rho}{(\cosh^2\rho+1)\sinh\rho}
\Big[e^{-\frac{i}{2}\rho\,\Gamma_\star\Gamma_1
+\frac{i}{2}\theta\,\Gamma_\star\Gamma_5}
\left(-\cosh\rho\,\Gamma_{01}+\Gamma_{05}\right)
\\&\hskip1in
+e^{\frac{i}{2}\rho\,\Gamma_\star\Gamma_1
-\frac{i}{2}\theta\,\Gamma_\star\Gamma_5}
\left(-\sin\theta\,\Gamma_{61}
+\sin^2\theta\,\Gamma_{65}\right)\Big]K\epsilon_0
\\
=&\ \frac{\cosh^2\rho}{(\cosh^2\rho+1)\sinh\rho}\,
e^{-\frac{i}{2}\rho\,\Gamma_\star\Gamma_1
-\frac{i}{2}\theta\,\Gamma_\star\Gamma_5}
\\&\ \times
\Big[
e^{i\theta\,\Gamma_\star\Gamma_5}
\left(-\cosh\rho\,\Gamma_{01}+\Gamma_{05}\right)
+e^{i\rho\,\Gamma_\star\Gamma_1}
\left(-\sin\theta\,\Gamma_{61}
+\sin^2\theta\,\Gamma_{65}\right)\Big]K\epsilon_0\,.
\end{aligned}
\end{equation}
Finally we expand the exponents in the last line, use 
(\ref{one-J-S-solution}) to relate $\theta$ and $\rho$ 
and the complex conjugate of (\ref{SUSY-cond-1}) to replace 
$\Gamma_6$ by other gamma matrices. Then almost all the terms cancel 
and we are left simply with
\begin{equation}
\Gamma\epsilon=
-e^{-\frac{i}{2}\rho\,\Gamma_\star\Gamma_1
-\frac{i}{2}\theta\,\Gamma_\star\Gamma_5}
\,\Gamma_{01}K\epsilon_0\,.
\end{equation}
So the projector equation $\Gamma\epsilon=\epsilon$ is solved by 
all constant spinors satisfying (\ref{SUSY-cond-1}) and
\begin{equation}
-\Gamma_{01}K\epsilon_0=\epsilon_0\,.
\label{SUSY-cond-2}
\end{equation}
It is easy to verify that those two conditions are consistent with each-other, 
so there are eight linearly independent real solutions to this equation. Thus 
the string solution preserves $1/4$ of the supersymmetries.

Note that each of those two conditions by themselves correspond to 
half-BPS configurations. (\ref{SUSY-cond-1}) relates the propagation 
in time to the rotation around a big circle on $S^5$, This condition is 
appropriate to the string state dual to the local operator $\Tr Z^J$ 
(the BMN ground state) which preserves half the supersymmetries of 
the $AdS$ background. For large $\sigma$ when $\rho\to0$ and 
$\theta\to\pi/2$ our solution approaches the BMN regime (as shall be 
explained in the following subsection) and 
that part of the world-sheet preserves half the supersymmetries.

On the other hand for small $\sigma$ (near the boundary of $AdS$)
\begin{equation}
\Gamma\sim-\Gamma_{01}K\,,
\end{equation}
so any constant spinor satisfying (\ref{SUSY-cond-2}) will solve the 
projector equation. That is not surprising, since near the boundary of 
$AdS$ the effect of the scalar insertions at past infinity are not noticed 
and the surface looks like that of the half-BPS straight Wilson line.

So in those two asymptotic regimes the surface preserves half the 
supersymmetries, but globally it preserves the intersection of the two 
conditions and is $1/4$ BPS.

\subsection{BMN limit}
Let us focus on the part of the world-sheet near the center of $AdS_5$. 
There we may go to the BMN limit \cite{Berenstein:2002jq}, by 
taking $L\to\infty$ and using the coordinates $x^+=(t+\phi)/2$, 
$x^-=L^2(t-\phi)/2$, $y=L(\pi/2-\theta)$ and $r=L\rho$. 
Combining $r$ and $y$ with the $S^3$ components of $AdS_5$ and 
$S^5$ into four-vectors, the metric becomes
\begin{equation}
ds^2=-4dx^+dx^--(\vec r\;^2+\vec y\;^2)(dx^+)^2
+d\vec r\;^2+d\vec y\;^2\,.
\end{equation}
The solution above survives in the limit and turns into
\begin{equation}
x^+=\tau\,,\qquad
y_1\sim|r_1|\,.
\end{equation}
To explain the last equality, $y_1$, which corresponds to the distance above 
the equator is always positive while $r_1$, which is the rescaled $\rho$ is 
allowed to extend to both positive and negative values, corresponding to 
the motion in the direction of two points on the boundary $S^3$.

Since this is a solution of the full theory, it will also be a solution in this 
limit. Consider the string Lagrangean in the light-cone gauge 
\cite{Metsaev:2001bj}
\begin{equation}
\cL=\frac{1}{4\pi\alpha'}\left[
(\dot{\vec r}\;)^2-(\vec r\;')^2
+(\dot{\vec y}\;)^2-(\vec y\;')^2-(\vec r\;^2+\vec y\;^2)\right]\,.
\end{equation}
Taking the ansatz $r_1=r_1(\sigma)$ and $y_1=y_1(\sigma)$ gives 
the equations of motion
\begin{equation}
r_1''-r_1=0\,,\qquad
y_1''-y_1=0\,,
\label{harmonic}
\end{equation}
and we choose the solution
\begin{equation}
r_1=a\sinh\sigma\,,\qquad
y_1=a\cosh\sigma\,.
\end{equation}
In the limit when $a\to0$ this reduces to a string with a right-angle 
$y_1=|r_1|$. The solutions with finite $a$ can also be extended to the 
full metric, if instead of $w_1=\kappa$ we take them slightly different, 
the surface will not get all the way to the equator, but will stay a 
bit north of it. After taking the BMN limit, this distance from the 
equator becomes $y_{\min}=a$.

Those are solutions of the full string equations of motion in Lorentzian 
signature, but the equations (\ref{harmonic}) are identical to those 
for a particle in an inverted 2-dimensional harmonic oscillator. The 
solution corresponds to a particle coming from infinity with energy 
very close to the maximum of the potential and with a small impact 
parameter, so it is deflected by a $\pi/2$ angle.

This solution naively carries infinite energy, but this is the standard 
divergence associated with the infinite string solutions describing 
Wilson loops \cite{Drukker:1999zq}. It does carry infinite angular 
momentum $J$. In the light-cone gauge the extent of the world-sheet 
coordinate $\sigma$ is identified with the conserved momentum 
$2\pi\alpha'p^+$ conjugate to $x^-$, related to the angular 
momentum by $J=L^2p^+=\alpha'\sqrt\lambda\,p^+$. The BMN 
background we are considering does not contain D-branes, so the 
strings have to extend to infinity and carry infinite $J$.

In order to study the small fluctuations around this solution we will 
need to impose a cutoff on $J$. A careful treatment will require 
quantizing the full solution (\ref{one-J-AdS-solution}), 
(\ref{one-J-S-solution}) beyond the BMN limit. But we 
expect the excitations to be confined close to the region described 
by the pp-wave metric, so we will try to quantize open strings with 
a finite $J$ in that limit ignoring the fact that the strings cannot end 
at finite $r$ and $y$.

Since the equations of motion are linear they are not affected by the 
background solution and will be solved by harmonic functions, like
\begin{equation}
y_i=e^{i\omega_p\tau}\sin p\sigma\,.
\end{equation}
The energy of such a state will be similar to the closed string 
excitations
\begin{equation}
\omega_p=\sqrt{1+p^2}\,.
\end{equation}
The only difference is the quantization condition on $p$, instead 
of requiring periodicity over the interval of 
$0\leq\sigma\leq 2\pi J/\sqrt\lambda$, 
we take functions that satisfy Dirichlet boundary conditions at the 
ends of such an interval. Hence the allowed values of $p$ are 
half of those for closed strings $p=n\sqrt\lambda/2J$ for 
integral $n$.

The dimension of such an excitation is therefore
\begin{equation}
\Delta-J=\omega_n=
\sqrt{1+\frac{\lambda n^2}{4J^2}}
\sim1+\frac{\lambda n^2}{8J^2}\,,
\end{equation}
This precisely agrees with (\ref{bethe-spectrum})!

The anomalous dimension is four times smaller than for the closed 
string excitations with $\lambda n^2/2J^2$. On both the spin-chain 
and the string calculations this factor of four comes from changing 
from functions that are periodic over the interval to functions whose 
period is double the interval.

\subsection{Solution with two angular momenta}
\label{2-J-section}

In the general case, when both $w_1$ and $w_2$ are non-zero 
the equations of motion are
\begin{equation}
\begin{gathered}
\rho''-\omega^2\cosh\rho\sinh\rho=0\,,\\
\theta''-\sin\theta\cos\theta
((\psi')^2-(w_2^2-w_1^2)\sin^2\psi-w_1^2)=0\,,\\
\psi''+\cot\theta\,\theta'\psi'
+(w_2^2-w_1^2)\sin\psi\cos\psi=0\,.
\end{gathered}
\end{equation}
The Virasoro constraint still gives two integrals of motion
\begin{equation}
-(\rho')^2+\omega^2\cosh^2\rho=
(\theta')^2+\sin^2\theta\left[(\psi')^2
+(w_2^2-w_1^2)\sin^2\psi+w_1^2\right]
=\kappa^2\,.
\label{virasoro2}
\end{equation}
Again we take $\omega=\kappa$ to get a solution that extends to all values 
of $\rho$ and get the same solution as before for the $AdS_5$ side
\begin{equation}
\sinh\rho=\frac{1}{\sinh\kappa\sigma}\,.
\end{equation}

On the $S^5$ side it's simple to check that the following quantity 
is also an integral of motion
\begin{equation}
\mu^2=\sin^2\theta\sin^2\psi
+\frac{\sin^4\theta(\psi')^2}{w_2^2-w_1^2}
+\frac{(\sin\psi\,\theta'+\sin\theta\cos\theta\cos\psi\,\psi')^2}
{w_2^2}\,.
\label{conserved-mu}
\end{equation}
On the boundary of the world-sheet, where $\sigma=0$ the string will 
sit at the point $\theta=0$. We should find a solution where $\psi$ is 
not a constant $0$ or $\pi/2$ so it carries angular momentum in both 
directions and in addition we have to require that the world-sheet extends 
to infinite $\sigma$ without an infinite number of oscillations.

To study this system it is useful to switch to ellipsoidal coordinate 
\cite{Babelon:1992rb,Arutyunov:2003uj} and find the solutions 
in terms of hyper-elliptic curves. 
We will not present that analysis here, since after studying that system 
of equations we found that it is possible to write the relevant solution 
easily in terms of $\theta$ and $\psi$. As it turns out, the requirement 
that the range of $\sigma$ diverge leads to a separation of scales. For 
small $\sigma$ the coordinate $\theta$ changes from $0$ to the 
final value of $\pi/2$. The coordinate $\psi$ changes at a much longer 
scale, remaining constant for all finite values of $\sigma$ and varying 
only on a diverging scale.

Let us solve the equations under those assumptions and then show that 
they are consistent. First take $\psi=\psi_0$ a constant, so $\psi'=0$, 
which leads to an equation for $\theta$ similar to the case with one 
angular momentum, (\ref{one-J-eom}). In this equation the terms 
proportional to $w_1^2$ and $w_2^2$ serve as potential terms, and 
the solution will correspond to a string coming in from infinity and 
climbing up to the top of the potential at an infinite time. Assuming 
$w_1^2<w_2^2$, the solution will go mainly in the 
less steep direction of $w_1^2$ at $\psi=0$. With the remaining 
residual energy it will then move away from $\psi=0$ into the other 
plane with rotation $w_2^2$ (the ellipsoidal coordinates mentioned 
above are useful to verify this).

Hence the Virasoro constraint is
\begin{equation}
(\theta')^2+w_1^2\sin^2\theta=\kappa^2\,.
\end{equation}
To get a solution that extends to infinite time we have to set 
$w_1^2\sim\kappa^2$ (the difference being infinitesimal, 
important in what follows) and the solution, as before, is
\begin{equation}
\sin\theta=\tanh\kappa\sigma\,.
\end{equation}

Now let us focus on the equations for $\psi$, which varies on a much 
longer scale than $\theta$ does, so we may now assume 
$\sin\theta=1$. Therefore we are studying the motion of the string inside 
an $S^3\subset S^5$ which is very similar to the system studied 
in \cite{Frolov:2003xy} where some classical ``folded-string'' 
solutions of the closed-string $\sigma$-model were described. 
Those solutions, which carry two angular momenta, have a profile that 
backtracks on itself to form a closed contour. In our case the solutions 
will be half of the folded strings; instead of closing on itself it will 
be connected to the part of the world-sheet with small $\sigma$ and 
extend to the boundary of $AdS_5$.

The second integral of motion (\ref{conserved-mu}) in this limit is simply
\begin{equation}
\mu^2=\sin^2\psi+\frac{(\psi')^2}{{w_2^2-w_1^2}}\,.
\end{equation}
This may immediately be solved by the elliptic integral of the first kind
\begin{equation}
\sigma\sqrt{w_2^2-w_1^2}
=\frac{1}{\mu}F\left(\psi,\frac{1}{\mu}\right)
=F\left(\arcsin\frac{\sin\psi}{\mu},\mu\right)\,.
\label{2J-solution}
\end{equation}
while $\psi$ varies from zero to its maximal value $\arcsin \mu$, the 
world-sheet coordinate will vary from $0$ to the complete elliptic integral
\begin{equation}
\sigma_\text{max}=\frac{K(\mu)}{\sqrt{w_2^2-w_1^2}}\,.
\label{sigma-max}
\end{equation}
If this interval is of finite length, the angle $\psi$ will oscillate an 
infinite number of times along the world-sheet. Therefore we must 
take $w_1^2\to w_2^2$, which allows for a finite number of 
oscillations (we take a single one). 
In this limit the variation of $\psi$ as a function 
of $\sigma$ vanishes, justifying our approximation above, which 
assumed $\psi'=0$ to solve for $\theta$.

Before proceeding with the calculation of the angular momentum 
carried by this solution, it's worthwhile pausing to explain the geometry 
of this peculiar solution. For finite values of $\sigma$, in most of 
$AdS_5$, the solution will look exactly like the the single angular 
momentum case, where $\psi=0$ and $\theta$ will increase from $0$ 
to $\pi/2$. Then for infinitely large $\sigma$, at the center of $AdS$, 
there will be another patch of world-sheet where $\theta=\pi/2$ and 
$\psi$ varies.

This second piece of the world-sheet is identical to half of a 
``folded-string'' solution of the closed-string $\sigma$-model. Instead 
of closing off on itself it is connected to the boundary of space to 
describe the Wilson loop observable. In order for the two regimes to 
connect smoothly in the conformal gauge, the part of the string where 
$\psi$ varies has to be rescaled by an infinite amount and will 
dominate in the calculation of the conserved charges. 
Then the agreement between the closed 
spin-chain in the thermodynamic limit and the ``folded-string'' 
solution \cite{Beisert:2003ea} will automatically extend to our 
case, as we show now.

The quantum number carried by the string will be given by integrals 
over the solution (\ref{2J-solution}) plus finite boundary terms from 
the region where $\sin\theta\neq1$ and $\cosh\rho\neq1$. The 
boundary terms are the same as for the single angular momentum case
(\ref{one-spin-J}) and (\ref{one-spin-E})
\begin{equation}
\begin{aligned}
J_1=&\ \frac{\sqrt{\lambda}}{\pi}
\left(\cJ_1-\tanh\kappa\sigma_\text{max}\right)\,,
\\
J_2=&\ \frac{\sqrt{\lambda}}{\pi}\cJ_2\,,
\\
E=&\ \frac{\sqrt{\lambda}}{\pi}
\left(\cE-\coth\kappa\sigma_\text{max}\right)\,,
\label{eq:J1-J2-E}
\end{aligned}
\end{equation}
where $\cJ_1$, $\cJ_2$ and $\cE$ are the expression derived from the 
solution (\ref{2J-solution}) and are essentially identical to the 
folded-string expressions \cite{Frolov:2003xy,Beisert:2003ea}%
\footnote{To be precise, in our case they are a $\pi/2$ of those in 
\cite{Frolov:2003xy,Beisert:2003ea}. That is due to the normalization 
in (\ref{eq:J1-J2-E}) and that the open string contains only half 
of the folded-string.}.
Those are given by complete elliptic integrals of the first and second kind
\begin{equation}
\begin{aligned}
\cJ_1=&\ \int_0^{\sigma_\text{max}}d\sigma\,
w_1\sin^2\theta(\sigma)\cos^2\psi(\sigma)
=w_1\sigma_\text{max}\frac{E(\mu)}{K(\mu)}\,,
\\
\cJ_2=&\ \int_0^{\sigma_\text{max}}d\sigma\,
w_2\sin^2\theta(\sigma)\sin^2\psi(\sigma)
=w_2\sigma_\text{max}\left(1-\frac{E(\mu)}{K(\mu)}\right)\,.
\\
\cE=&\ \kappa\sigma_\text{max}\,.
\end{aligned}
\end{equation}
where for now we keep the $\sigma_\text{max}$ finite.

These two equations together with the expressions for $\sigma_\text{max}$ 
(\ref{sigma-max}) and the relation 
$\kappa^2=w_1^2+(w_2^2-w_1^2)\mu^2$ may be summarized by
\begin{equation}
\frac{\cE^2}{K(\mu)^2}
-\frac{\cJ_1^2}{E(\mu)^2}
=\mu^2 \,,
\qquad
\frac{\cJ_2^2}{(K(\mu)-E(\mu))^2}-\frac{\cJ_1^2}{E(\mu)^2}
=1\,.
\end{equation}
These are the same equations as those of the folded $2$-spin
solution \cite{Frolov:2003xy,Beisert:2003ea}.

with $\cJ=\cJ_1+\cJ_2$ we may expand these expressions at large 
$\cJ$ to find
\begin{align}
\frac{\cJ_2}{\cJ}\sim&\ 1- \frac{E(\mu)}{K(\mu)} \,,
\\
\cE\sim&\ \cJ+\frac{1}{2\cJ}
K(\mu) \left(E(\mu) - (1-\mu^2) K(\mu) \right) \,.
\end{align}
Using those expressions we can go back to the full solution including 
the boundary effects (\ref{eq:J1-J2-E}). In the limit of large 
$\kappa\sigma_\text{max}$ we find
\begin{align}
\frac{J_2}{J}\sim&\ 1- \frac{E(\mu)}{K(\mu)} \,,
\\
E\sim&\ J+\frac{\lambda}{2\pi^2J}
K(\mu) \left(E(\mu) - (1-\mu^2) K(\mu) \right) \,.
\end{align}

To compare those expressions to (\ref{elliptic}) we define
\begin{equation}
\mu=i\frac{1-\sqrt{1-k^2}}{2(1-k^2)^{1/4}}\,,
\end{equation}
and use the modular transformation of the elliptic integrals 
\cite{Beisert:2003ea}
\begin{equation}
K(\mu)=(1-k^2)^{1/4}K(k)\,,\qquad
2E(\mu)=(1-k^2)^{-1/4}E(k)+(1-k^2)^{1/4}K(k)\,,
\end{equation}
to find the relations
\begin{align}
\frac{J_2}{J}\sim&\ 
\frac{1}{2}-\frac{1}{2\sqrt{1-k^2}}\frac{E(k)}{K(k)} \,,
\\
E\sim&\ J+\frac{\lambda}{8\pi^2J}
K(k)\left(2E(k)-(2-k^2)K(k)\right) \,.
\end{align}
This is exactly the same as the results of the Bethe equations in the 
thermodynamic limit (\ref{elliptic}) under the replacements 
$J\leftrightarrow K$, $J_2\leftrightarrow M$
and $\gamma_M\leftrightarrow E-J$.

This agreement is a direct consequence of the relation between the folded 
strings and the two-cut solution of the usual closed $SU(2)$ spin-chain. 
On the spin-chain side the open string is described by a single cut, the 
other cut is an image and all the charges are half as in the close spin-chain 
case. The string theory solution is just half of the folded string connected 
to the boundary of space and the extra boundary terms did not change the 
relations between the charges at this order in the expansion in $\lambda/J$. 
We also expect that the fluctuations 
of this open string, as in the example of the single spin, will be confined 
to the middle of the string and not get close to the boundary, so their 
spectrum will be the same (up to a numerical factor $1/4$) to the 
closed-string analog.

For closed strings a general classification of classical solutions carrying 
two charges was provided in \cite{Kazakov:2004qf} which completely 
agrees with possible solutions of the $SU(2)$ Bethe ansatz in the 
thermodynamic limit. One may hope that a similar analysis will be valid 
in our case. All the solutions of the open spin-chain thermodynamic 
Bethe ansatz can be related to solutions of the closed system invariant 
under the symmetry $u\to-u$ for the roots. Those should all be 
described by folded string solutions in the dual string theory. It seems 
quite plausible to connect half of those strings to the boundary in 
a similar fashion to the case studied above and thus describe more general 
Wilson loop observables.

\section{Correlator with a local operator}
\label{sec-local}

Up to now we studied Wilson loops with two insertions of operators 
in the adjoint. We may also look at the two point function between 
a loop with a single insertion and a local operator. The case of the 
two point function of the loop with no insertion to a local operator 
made up of $\Phi_6$ was considered in the past 
\cite{Semenoff:2001xp,Pestun:2002mr}
In that case agreement was found between the perturbative calculation 
and that from $AdS_5\times S^5$.

Consider first a straight line with an 
insertion at the origin of a word made up of $Z$s and $X$s. Then take 
another single-trace local operator at time $t$ and a distance $r$ from the 
line. Since 
the straight line as well as the origin are invariant under dilatation, we 
can use the Ward identity associated with the broken dilatation symmetry 
to get the partial differential equation
\begin{equation}
(r\,\partial_r+t\,\partial_t+\Delta+\Delta')
\vev{\Tr\bar\cO'(t,r)\, W[\cO(0)]}=0\,.
\end{equation}
The general solution to this equation takes the form
\begin{equation}
\vev{\Tr\bar\cO'(t,r)\, W[\cO(0)]}
=\frac{f(t/r)}{(t^2+r^2)^{(\Delta+\Delta')/2}}\,,
\end{equation}
where $f(t/r)$ is an arbitrary function.

Due to the extra arbitrary function one cannot automatically associate 
all logarithmic divergences with normalization of the conformal 
dimension $\Delta$. Doing that, and fixing the function $f$ requires 
more work. Instead we will consider the case of $r=0$, when the 
local operator is along the line, and then $f$ is just a constant, which 
may depend on the coupling, but not on $t$. Then indeed all 
logarithmic divergences are associated to the renormalization 
of $\Delta$.

In the case of the circle, if the local operator is located at the 
position given by $\rho$, $\psi$ and $\tilde r$ defined in 
(\ref{coordinates}), the Ward identity associated with $J_0$ is
\begin{equation}
\tilde r^{-\Delta'}(-\cos\psi\,\partial_\rho
+\coth\rho\sin\psi\,\partial_\psi+\Delta)\tilde r^{\Delta'}
\vev{\Tr\bar\cO'(\rho,\psi)\, W[\cO(\infty,0)]}=0\,.
\end{equation}
Note that the circle is at $\rho\to\infty$. 
The general solution to this equation takes the form
\begin{equation}
\vev{\Tr\bar\cO'(\rho,\psi)\, W[\cO(\infty,0)]}
=\frac{f(\sin\psi\sinh\rho)}
{(2R(\cosh\rho-\sinh\rho\cos\psi))^\Delta\tilde r^{\Delta'}}\,.
\end{equation}
For $\Delta'=\Delta$ the function in the denominator is the distance 
from the local operator to the insertion at $\psi=0$ along the Wilson 
loop to the power $2\Delta$. Again $f$ is an arbitrary function, and 
we will restrict to the case where $\rho\to\infty$, or the local operator 
sits on the circle at radius $R$ to eliminate that ambiguity.

At tree-level the Wilson loop will reduce to the trace of a local operator, 
and the two point function will be zero unless $\cO$ and $\cO'$ are 
identical up to cyclic transformation. At one loop if we consider the 
diagrams that do not involve the Wilson loop, those again will be the same 
as for two local operators. In addition there are graphs where the holonomy 
is expanded to first or second order. The latter is finite, but the first 
has divergences.

So at the one-loop level these extra graphs will be the only difference 
from the system of two single-trace local operators. Those graphs are 
the same as calculated in Appendix~\ref{sec-calc} and discussed 
in Section~\ref{sec-one-loop}. But while at each end of the integration 
region there is a divergence, a careful accounting of the signs reveals that 
their sum vanishes.

Thus this system is described by the usual Heisenberg spin-chain in the 
$SU(2)$ sector
\begin{equation}
H_2=\frac{\lambda}{8\pi^2}\sum_{k=1}^K(I-P_{k,k+1})\,.
\end{equation}
From this one-loop calculation one cannot see the difference between a local 
operator described by a closed spin-chain and the Wilson loop, related to 
an open one. We expect that at higher levels 
in perturbation theory, or when considering insertions of other fields 
the situation will be more complicated. The Hamiltonian may contain 
more terms localized near the endpoints of the insertion at $k=1$ and 
$k=K$.

In the general case the system will be described by a closed loop with a marked 
point. This marked point will provide all the extra interactions associated 
with the Wilson loop and the impurities along the spin-chain which may be 
transmitted through it (as in the one-loop calculation above) or reflected 
from it with certain amplitudes.

\section{Discussion}

The main idea of this paper is to consider small deformations of 
circular/straight Wilson loops and study them using conformal 
field theory techniques. Local deformations of the loop are analogous 
to insertions of adjoint operators into the loop and we concentrated 
on insertions of words made of two complex scalar fields.

Since the Wilson loop without insertions preserves part of the 
superconformal group which includes a factor of $SL(2,\bR)$, we 
may classify local insertions in terms of representations of that 
group. Thus we associate a conformal dimension to operators 
transforming in the adjoint representation. 
Calculating the dimensions of the scalar insertions we were 
driven to study open spin-chains with very simple boundary 
conditions. We presented the one-loop anomalous dimensions 
of those insertions both in the dilute gas approximation and the 
thermodynamic limit by a simple modification of the standard 
results for closed spin-chains.

We then went over to study the system in string theory, where the 
Wilson loop is given by a classical string surface in 
$AdS_5\times S^5$. We found the relevant surfaces and calculated 
the anomalous dimensions of those operators at strong coupling. 
Again we found that the string solutions is very similar to the ones 
for closed strings. Quite remarkably, it is possible to take half of the 
closed ``folded-string'' solutions and instead of closing them on 
themselves, extend them to the boundary of space in order to describe 
our Wilson loop observables. The dimensions we calculated 
in this way agree with the perturbative results.

Clearly it would be interesting to study this system further in the 
usual ways: Go beyond 1-loop to higher loop amplitudes, compare 
with the Hubbard model \cite{Rej:2005qt} and include a wider class 
of insertions involving the other scalars, fermions and 
field-strengths. One may also look at multiply wrapped Wilson 
loops as well as 't Hooft loops. We expect those to yield interesting 
open spin-chains that are definitely worth exploring.

Note that in all the discussion here the subtle difference between the 
line and the circle \cite{Drukker:2000rr} did not play any role. 
The anomalous dimension comes from 
divergences in the loop that happen at short distances and therefore 
are not affected by this global issue. If we were to calculate 
finite terms like the proportionality constant $C$ in 
$\vev{W[\cO'(t)\,\cO(0)]}=C/t^{2\Delta}$, this constant cannot 
depend on the distance but it can be a function of the coupling 
and may be different for the line and the circle.

Our main motivation is not finding spin-chains, those are merely a 
fascinating calculational tool. It is studying Wilson loop operators, 
some of the most interesting observables in gauge theories. Instead 
of studying the most general operators following an arbitrary path 
we focused on small deformations of the circular/straight operator. 
In principle it is possible to build back an arbitrary path from many 
such insertions, but it will require a lot more work.

We see here that it is possible to treat the Wilson loop in a similar 
fashion to a defect CFT 
\cite{Cardy:1984bb,McAvity:1995zd,DeWolfe:2001pq}. 
Those are theories with extra degrees of freedom living on a 
sub-manifold, but still preserving part of the conformal group. Here 
though, the Wilson loop and the insertions that we attach to it are 
already an integral part of the theory and one does not have to 
introduce extra degrees of freedom. The most novel thing about 
this CFT living on the Wilson loop is that it allows operators that are 
not singlets of the gauge group!

\section*{Acknowledgments}
We are grateful to Gleb Arutyunov, Niklas Beisert, Bartomeu Fiol, 
Lisa Freyhult, Sunny Itzhaki, Yuri Makeenko, Matthias Staudacher and 
Arkady Tseytlin for interesting conversations. N.D. would like to thank 
Tel Aviv University for its hospitality at the final stages of this work. 
The work of S.~K. was supported by ENRAGE (European Network 
on Random Geometry), a Marie Curie Research Training Network 
supported by the European Community's Sixth Framework Programme, 
network contract MRTN-CT-2004-005616.

\appendix

\section{Interaction of a local insertion with the Wilson loop}
\label{sec-calc}

We show here the details of the calculation of the Feynman graph 
depicted in figure~\ref{boundary-fig}. This graph gives the boundary 
term for the spin chain. The result can be extracted from the calculations 
of Erickson, Semenoff and Zarembo \cite{Erickson:2000af} and we 
follow their conventions.

It is enough to consider the correlator of a single $Z$ and a 
single $\bar Z$ insertion into the straight Wilson loop. The relevant 
graph is the exchange of a gluon between the scalar propagator and 
the Wilson loop.

Let us take one scalar at the origin ($t_1$), another at $t_3=t>t_1$ 
and a gauge field (from the expansion of the Wilson loop) at $t_2$ 
between them. There will be another graph when $t_2$ is outside of 
this interval, doubling the final answer. This graph is equal to
\begin{equation}
\frac{1}{N}\int_{t_1}^{t_3}dt_2\,
\vev{\Tr [\bar Z(t_3)iA_t(t_2)Z(t_1)]}\,.
\end{equation}
Here $Z(t)=Z^a(t)T^a$, where $T^a$ are the generators of $U(N)$
obeying
\begin{equation}
\sum_aT^aT^a=\frac{N}{2}{\bf1}\,.
\end{equation}
We contract this with the vertex
\begin{equation}
-\frac{1}{g_{YM}^2}\int d^4w\, f^{abc}\left(
\partial_\mu Z^a(w)A_\mu^b(w)\bar Z^c(w)+\hbox{c.c.}\right).
\end{equation}

After contracting $Z$ with $\bar Z$ and the gauge fields with the term 
written explicitly above in the vertex, the trace gives a factor of
\begin{equation}
-\frac{1}{g_{YM}^2N}\Tr(T^aT^bT^c)f^{abc}
=-i\frac{N^2}{4g_{YM}^2}\,.
\end{equation}
This term has a derivative acting on the $w$ coordinate in the $w$-$t_3$ 
propagator, or a $(-\partial_3)$ derivative. The complex conjugate 
term in the vertex gives a similar term with $\partial_1$.

So this graph contributes
\begin{equation}
\frac{N^2}{4g_{YM}^2}\int dt_2\int d^4w\,(\partial_1-\partial_3)
G(w-t_1)G(w-t_2)G(w-t_3)\,.
\end{equation}
where $G$ is the scalar propagator. This integral will diverge when 
$w\sim t_2\sim t_1$ and when $w\sim t_2\sim t_3$. 

Near the origin we may replace $G(t_3-w)=G(t_3-t_1)$ and then the 
integral over $w$ will give a result that depends only on $t_2-t_1$ and 
from dimensional grounds $\ln(t_2-t_1)$. More precisely
\begin{equation}
\int d^4w\,G(t_2-w)G(t_1-w)
=-\frac{g_{YM}^4}{8\pi^2}\ln(t_2-t_1)\,,
\end{equation}
where an IR cutoff has to be included to make the log well defined.

The $\partial_3$ derivative gives a finite answer, but the $\partial_1$ 
not. It gives a total derivative with respect to the $t_2$ integral, which 
we cutoff at $t_2-t_1=\epsilon\sim1/\Lambda$, so the final result is
\begin{equation}
-\frac{\lambda^2}{2^7\pi^4t^2}\ln\epsilon
=\frac{\lambda^2}{2^7\pi^4t^2}\ln\Lambda\,.
\end{equation}
Recall that the tree-level result was $\lambda/8\pi^2 t^2$, so to get a 
finite answer we have to renormalize the operator by 
\begin{equation}
Z_{\hbox{\scriptsize boundary}}
=I-4\frac{\lambda}{16\pi^2}\ln\Lambda\,.
\end{equation}
The factor of 4 comes from the two limits of the above integral as 
well as the other graph with $t_2>t_3$ and $t_2<t_1$. This boundary 
$Z$-term exactly cancels that from the self-energy correction, as was 
already observed in \cite{Erickson:2000af}. Note that we will 
associate half of it with each of the local insertions into the loop, 
giving (\ref{z-boundary}).


\end{document}